# Embedding Climate Change Engagement in Astronomy Education and Research

Authors: K. Williamson[1], T.A. Rector[2], J. Lowenthal[3],
[1]West Virginia University, [2]University of Alaska Anchorage, [3]Smith College
Contact: kewilliamson@mail.wvu.edu

Endorsed by the AAS Sustainability Committee, including C. De Pree[4], K. Olsen[5], A. Soto[6], N.A. Miller[7], E. Perlman[8], G. C. Clayton[9]
[4]Agnes Scott College, [5]National Optical Astronomy Observatory, [6]Southwest Research Institute, [7]Stevenson University, [8]Florida Institute of Technology, [9]Louisiana State University

Endorsed by an additional 323 astronomy professionals via a Change.org petition available at tinyurl.com/Climate-White-Paper-Signatures.

This White Paper is a call to action for astronomers to respond to climate change with a large-scale structural transition within our profession. Many astronomers are deeply concerned about climate change and act upon it in their personal and professional lives, and many organizations within astronomy have incorporated incremental changes. We need a collective impact model to better network and grow our efforts so that we can achieve results that are on the scale appropriate to address climate change at the necessary level indicated by scientific research; e.g., becoming carbon neutral by 2050. In particular we need to implement strategies within two primary drivers of our field: (1) **Education and Outreach**, and (2) **Research Practices and Infrastructure**. (1) In the classroom and through public talks, astronomers reach a large audience. Astronomy is closely connected to the science of climate change, and it is arguably the most important topic we include in our curriculum. Due to misinformation and disinformation, climate change communication is different than for other areas of science. Bad climate communication strategies can be ineffective and even backfire. We therefore need to expand our climate change communication and implement effective strategies, for which there is now a considerable body of research. (2) On a per-person basis astronomers have an outsized carbon impact. There are numerous ways we can reduce our footprint; e.g., in the design and operation of telescope facilities and in the optimization and reduction of travel. Fortunately, many of these solutions are win-win scenarios, e.g., increasing the online presence of conferences will reduce the carbon footprint while increasing participation, especially for astronomers working with fewer financial resources. Considering the size and severity of the problem, astronomers have an obligation to act on climate change in every way possible, and we need to do it now. In this White Paper, we outline a plan for collective impact using a Networked Improvement Community (NIC) approach.

# Introduction

From a social, environmental, and political standpoint, climate change is arguably the most important topic of our time. The need to address climate change has become immediate, with only 12 years to reduce carbon emissions to 50% of pre-1990's levels in order to avoid potentially irreversible and catastrophic effects (Intergovernmental Panel on Climate Change Special Report 2018). We are now in a crisis scenario, and we need all hands on deck to address this problem. While many astronomers recognize the urgency of addressing climate change and may already take personal actions, this White Paper highlights how and why astronomers should network together to coordinate our efforts.

*Climate change threatens astronomy*. For example, we build telescopes in dry and stable climates, such as the Atacama desert. But the climatic stability of these locations is now threatened by extreme weather events and changing weather patterns. Suggested geoengineering strategies, such as releasing aerosols into the atmosphere to reduce solar flux, are just as threatening if not worse. They would have dire consequences for astronomy by hampering our ability to use ground-based optical-IR telescopes. One recent estimate found that the level of stratospheric aerosol injection (SAI) required to achieve useful reduction in incoming solar radiation would result in not only a 25% increase in urban night sky brightness, but also a factor of two increase in scattering of starlight in the atmosphere (Zender 2016). Clearly, the best solution is to stop carbon emissions rather than reduce solar flux, and this can only be done if we act now.

*Our professional practice creates a large carbon footprint*. We build large facilities and travel extensively. In the previous decadal review, a white paper by Marshall et al. (2009) outlined strategies for astronomers to reduce our impact. In the ten years that have since passed, we have learned more about the severity of climate change and the need for urgent action. Many organizations within astronomy have already implemented policies on a range of scales, e.g., remote observing; the elimination of single-use water bottles at Las Campanas Observatory, the installation of solar panels at Gemini South, and a greater use of remote conferencing for panel reviews at the NSF. Other science organizations, such as the American Geophysical Union and the American Chemical Society, have also stepped up their efforts by releasing statements and developing action plans.

*Astronomers are well-poised to communicate climate change to others*. Recent surveys in 2018 by the Yale Program on Climate Change Communication show that the majority of Americans are concerned about climate change, but their perception and understanding of climate change is far behind the scientific consensus. Furthermore, only 38% of Americans talk about climate change occasionally, and only 22% hear about it in the news on a weekly basis. As a result it is not seen as important. Simply put, people need to hear about climate change more frequently, and in a variety of contexts. In recent years we have learned that the traditional methods for science communication, e.g., the impartial presentation of facts, generally doesn't motivate people to act. In fact, bad communication can actually make the problem worse. Fortunately there is now a substantial body of climate communication research that we can use to be better educators. The major challenge is not teaching climate science content knowledge, but overcoming misinformation and disinformation to increase awareness of the urgency of the problem such that people will be motivated to act.

*We don't have excuses.* As noted by climate scientist Dr. Katharine Hayhoe, the most important thing we can do to fight climate change is to talk about it.  Historically scientists have been notoriously reluctant to engage in advocacy. Fears of losing credibility, "politicizing science," or of retaliation are often cited by scientists as reasons not to advocate, but a recent study by Kotcher et al. (2017) indicates that these fears are unfounded. Furthermore, astronomers are in general highly regarded, and in some ways more trusted than other branches of science because our research is less controversial (e.g., Smith 2003).  We reach large audiences through our college introductory courses, public talks, and other outreach.  And, we are less likely to be accused of being "in it for the money" because very few astronomers receive funding for climate science. Indeed, it is helpful for people to hear about climate change from many different kinds of scientists, because climate change is connected to all branches of science.

With this White Paper, we hope to catalyze a culture within our  discipline that includes regular and vigorous conversations on climate change, as well as immediate action. We may not all agree on how to move forward, but we believe that by normalizing the conversation we can build a shared vision of values and of hope, not only for a better profession, but for a better world.

## A Networked Improvement Community Approach

How do we position our entire discipline to embed climate change engagement in the way we do business? To conceptualize such a large transformation, we may turn to the theories of Networked Communities and Improvement Science, or "Networked Improvement Communities (NICs)," originally developed by the Carnegie Foundation for the Advancement of Teaching (CFAT). NICs have roots embedded in rapid prototyping methods of engineering design and have been used in educational settings and even climate literacy efforts, such as through the CLEAN Network (Ledley et al. 2014). We use the NIC model to structure this White Paper because it honors the complexity of a problem and empowers people within a community to take collective action, and this approach is gaining traction in large efforts nationwide. As scientists, we are already familiar with designing and testing our ideas, coordinating with others, and linking many small actions to achieve big results. We can apply this same logic to our aim of tackling climate change as a community of astronomers.

Taking a Networked Improvement Community approach aligns us with broader efforts to improve STEM and STEM Education happening nationwide, such as through "NSF's 10 Big Ideas." The NSF INCLUDES Alliance effort, in particular, has adopted the NIC approach for each of their five major Alliance-funded projects (NSF News Release 18-074). For example, the First 2 Network (of which Williamson is a part) uses NIC to help rural, first-generation college STEM majors persist through their first two years of college through implementing "change ideas" such as early research experiences (i.e. during the summer before their Freshman year) and a Hometown Ambassador program, in which students reach out to younger students from their high schools. The NIC approach, therefore, has broad applicability, and it has great potential to open doors for astronomers to new partnerships and funding opportunities.

In an effort to start the conversation for what a NIC approach might look like, we provide a "Driver Diagram" in Figure 1 below. The "primary drivers" of our discipline are (1) Education and Outreach, and (2) Research Practices and Infrastructure. These are the big umbrellas under which we do business. Every astronomer works under these umbrellas, but we may differ in how we interact with each area. The smaller spheres of influence where we may find ourselves doing work are the "secondary drivers." For example, "Astro 101" represents only one space where astronomy education happens. Planetaria and outreach programs represent others. Finally, within each of these secondary drivers are small, manageable "change ideas" for what we do with our "boots on the ground." For example, in teaching Astro 101, a professor can choose to devote more lessons to the greenhouse effect and global warming, assign readings to students, and empower them to use their own voices by writing letters to decision makers.

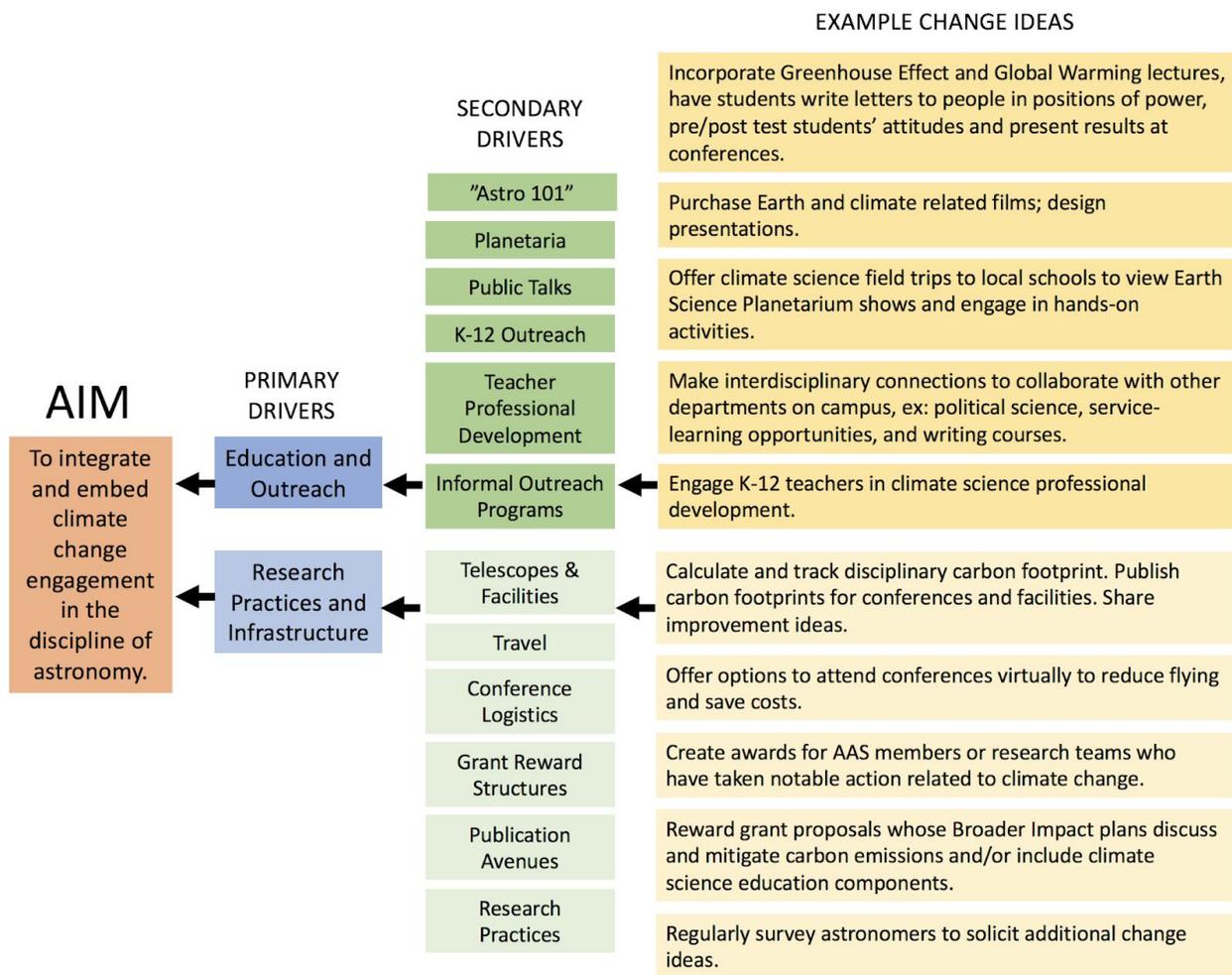

Figure 1: Driver Diagram demonstrating how astronomers can embed climate change engagement in our professional work.

The Driver Diagram in Figure 1 is by no means complete, and the yellow "change ideas" shown are meant only to be examples of concrete actions we might take as a community to effect change. If we move to adopt and formalize this approach, we would likely want to include NIC workshops in our conferences, in which large numbers and multiple groups of astronomers have the opportunity to contribute to characterizing the problems and solutions. Perhaps these workshops could be facilitated by NIC experts from the Carnegie Foundation.

In the following sections we discuss how and why the NIC approach can propel us forward through each of our Primary Drivers of (1) Education and Outreach and (2) Research Practices and Infrastructure.

## Education and Outreach

*We need to create a disciplinary culture in which effective climate change education is integrated as standard practice in all of our educational spaces.*

Communication is key. Climate science is closely related to astronomy; and most astronomers already know enough about the basics to teach it. However, the politicization of climate science means we need to approach the topic differently than others. In fact, poorly executed climate science communication can make the problem worse, causing doubters to become more deeply entrenched in their beliefs. Climate change communication research has identified strategies that are more effective in changing attitudes. These include the use of storytelling instead of informational narratives (e.g., Morris et al. 2019), focusing on issues relevant to the viewing audience; i.e., keeping it "here and now" by focusing on local impacts (e.g., Lorenzoni et al. 2007), and addressing the roles that individual and cultural identities play in decision making (e.g., Ferguson et al. 2016). Resources such as the Alan Alda Center for Communicating Science, including their Communicating Climate Change workshop, offer useful professional development opportunities.

Teaching climate change in "Astro 101" is particularly important because these classes contain a large number of students. Roughly a quarter million people enroll in an introductory astronomy class every year (Fraknoi 2001). Many undergraduate students have misconceptions about the causes of climate change, or if it is indeed real (e.g., Thiessen 2011). For many students, Astro 101 is their last formal education in the sciences, making this their best chance to learn the science behind climate change. Since Astro 101 is intended for non-scientists, it reaches an important demographic, e.g., voters, business leaders, and especially future teachers.

Faculty should include more climate science lessons, a natural extension for astronomy because course topics overlap with climate science topics (e.g. the greenhouse effect). A survey question in a recent Astro 101 class taught by one author (Williamson) asked how these lessons impacted students. Responses included, "I apparently didn't even know what the greenhouse effect or global warming really was until this class. I think that growing up, we just hear these terms so often with the simplification of 'It means that people are making the planet warmer,' that it's simply embarrassing to have to be told what it's really all about." In addition to including more lessons on climate science, faculty can connect to broader university goals (e.g. service learning and community engagement) by having climate scientists and activists come to class as guest speakers. Students can practice advocacy skills by writing letters to "someone in a position of power" (sending the letters can be optional). This practice has the benefit of moving away from "doom and gloom" lessons and can be transformational for empowering students to take action.

To formalize discussions about how to better incorporate climate change in Astro 101, one of us (Rector) hosted a workshop at the AAS Winter 2019 meeting in Seattle, which 70 astronomers attended. This attendance reflects the strong interest within our community to engage in these discussions, and we can capitalize on this momentum through the NIC model, e.g., scheduling regular climate change sessions and/or workshops at every AAS meeting. In keeping with a scholarly approach, we foresee the need to incorporate astronomy education research (AER) on the effectiveness of these professional development and educational strategies.

Astronomers are also leaders in public outreach, such as through planetaria and public talks. Planetaria alone reach over 146 million people worldwide (Petersen 2018). The NASA Data Visualization Studio has stunning data-driven videos of Earth systems that can be freely downloaded and used in any setting, with planetarium domes being particularly inspiring. Several of these visuals were incorporated into the film, "Dynamic Earth: Exploring Earth's Climate Engine," which now screens at over 175 planetariums worldwide. Additionally, as part of their professional service, astronomers participate in tens of thousands of public talks and outreach events that reach other vitally important educational spaces, such as national parks and public libraries. For example, since 1996 the "Universe in the Park" project spearheaded in Wisconsin has taken portable telescopes to state parks and engaged visitors in short presentations and answering questions. More recently, the "Astronomy on Tap" informal lecture series has become popular, with over 150 events at dozens of locations (Rice & Levine 2016), in which astronomers give low-key lectures at local bars and answer questions. Our immersion as astronomers in these public spaces offers critical entry points for connecting with the general public. We should use these opportunities to incorporate some discussion of climate change into what we are already doing, and doing well.

## Research Practices and Infrastructure

*We need to create a disciplinary culture in which minimizing our carbon footprint is valued, incentivized, and rewarded as a high priority.*

We need to have regular conference sessions, working groups, and collaboration spaces for understanding how climate change impacts astronomy. For example, at the January 2017 meeting in Grapevine, TX, the AAS Sustainability Committee hosted a session on "Geoengineering the Atmosphere to Fight Climate Change: Should Astronomers Worry About It?" attended by approximately 100 astronomers. We have also had other well-attended sessions including role-play exercises on how to have conversations about climate change, and keynote talks by top climate scientists, including authors of IPCC reports. We should work to standardize these practices and showcase our commitment on our web presence. For example, both the American Chemical Society and the American Geophysical Union have a variety of online climate change resources. Perhaps the thrust for these efforts can be accomplished through significant expansion of the Sustainability Committee, i.e. more financial resources, representation, leadership, and power to influence the AAS strategic plan and vision. This is well aligned with the NIC need for a "backbone" group to oversee and coordinate efforts across a network.

Second, we need to track our carbon footprint and travel less. Individuals can track their professional carbon footprints, and organizations such as observatories can track the number of users they serve to report carbon output per user. Travel is a problem in academia in general, accounting for as much as one third of a university's carbon footprint (Hiltner 2016). But in addition to conferences, panel reviews, and invited talks, astronomers may have an outsized contribution because of travel to remote telescope facilities worldwide. Thanks to internet connectivity, we now have widespread use of remote observing. Queue and service observing, as well as robotic operations, are part of the standard operations of a number of facilities. There is also a growing role for surveys in astronomy, which may reduce overall travel for collecting data. Archival reuse of data may then become a significant part of astronomers' research efforts, decreasing the amount of effort they put into smaller PI-led observing programs. Additionally, many panel reviews (e.g. NSF) are now being done by remote video conferencing, and we make great use of communication and collaboration tools such as Slack, Zoom, and Google. But we can do better. For example, the Nearly Carbon Neutral conference model (Hiltner 2016) used in other fields structures conferences such that participants do not meet in person at all, but instead pre-record and upload their talks to a conference website and then contribute together to an interactive Q&A session. The government shutdown prior to and during the AAS Winter 2019 meeting in Seattle served as a good example of how well astronomers can adapt to this model. The planning committee successfully scrambled to quickly offer virtual posters and live streaming of many conference presentations to accommodate interrupted travel schedules. We can grow from this experience and explore possibilities to do this for future conferences. For example, the American Chemical Society offers Beam Pro Telepresence robots, in which a participant can maneuver a robot and live video screen to any conference location. Additional benefits of remote conferencing include: (1) reduced costs; i.e., increases in conference expenses for IT infrastructure will be more than offset by saving in travel and hotel costs, (2) increased productivity; i.e. less time in transit means more time for science, and (3) greater inclusivity; young professionals, those with families, and astronomers at smaller institutions may not have the financial means or time to travel to far-away conferences, so virtual conferences have great potential to reach underrepresented groups. Clearly, reducing our carbon footprint is a win for the environment and a win for astronomers.

Linking change to reward will be critical in moving forward. For example, as a simple step, the AAS could create a yearly Sustainability Award to be given to the researchers or research teams who have taken notable actions to reduce their carbon footprint and/or to incorporate climate change education in their work. Perhaps these awards could even be tied to funding incentives. Indeed, over the past few decades, the merit of proposals has been determined by an evolving list of factors related to Broader Impacts. Funding agencies reward proposals that present strong arguments for the importance of their research in impacting society, training future researchers, and educating others, etc. More recently, projects have been asked to include diversity and inclusivity measures. We can draw on this long tradition to require or reward projects that include an estimate of their carbon footprint, climate mitigation and sustainability strategies, and climate education goals.

# Recommendations

In this White Paper, we have outlined a Networked Improvement Community approach as a potential framework for embedding climate change engagement in the astronomy community through both Education and Research. We have proposed possible change ideas that may be implemented at the person-level and institution-level. These change ideas should be viewed as examples only. We would not want readers to reject the underlying premise or resist committing to taking action just because they disagree with one of the ideas presented here. Regardless of what form this action takes, we strongly advocate for formalizing collective impact in astronomy to continue these conversations and take concrete, rapid, effective action to combat global climate change.

*Individual impact*: Collective impact starts with the individual, so we summarize change idea recommendations for individual astronomers to take action and network with others.

- Increase the amount of time you talk about climate change, inside the classroom and in public outreach settings.
- Increase your knowledge of climate science, especially in areas pertinent to local impacts.
- Learn effective climate change communication strategies. Attend science communication workshops, and read the relevant literature.
- Implement ways to reduce your own carbon footprint. Track and reduce energy use in your home and office. Think carefully about the necessity of flying.
- Take a scholarly approach to understanding your effectiveness in education and outreach. Survey participants and use educational data to make pedagogical decisions.
- Publish and/or present what strategies you have tried and your lessons learned to other astronomers through conferences, journals, social media, blogs, letters, etc.
- Contribute to the conversation. Talk to colleagues at your institution to discuss ways your department can change. Speak up if no one in the room is talking about climate change. Ask questions about how climate change is being considered.
- Consider taking climate leadership roles in your department or organization, such as joining the AAS Sustainability Committee.

*Collective impact*: Networked improvement will be much more effective if we can make structural changes to the way we do business. Below are examples of ways observatories, labs, astronomy departments, and professional organizations such as the AAS can address climate change.

- Work to characterize their carbon footprints, track them each year, and aim for measurable reduction within rapid timescales. Maintain data archives to track longitudinally and compare with others.
- Publish climate change resources, statements, and endorsements on websites, flyers, and other dissemination avenues. Find out about and publicize carbon footprint/sustainability issues at the institutional level (e.g., what is your institution doing?).
- Articulate ways they can reduce travel for themselves and activity participants.

- Include climate advocacy, communication, and NIC training workshops and provide brainstorming and collaboration space for attendees to communicate about their change ideas (both in person and virtually).
- Regularly survey members to track perceptions of current efforts and extent of networked improvement. Clearly and frequently communicate results.
- Identify mutually reinforcing change ideas and shared systems for measuring impact.
- Conference organizers can experiment with Nearly Carbon Neutral conference models.
- Funding agencies should reward and incentivize activities and projects that incorporate climate change education and mitigation practices in their proposals.
- Organizations should consider rewarding members for effective climate change education and mitigation projects with new prizes and/or other recognition.
- Expand advisory panels and sustainability committees to hold others accountable for embedding effective climate change engagement in education and research spaces.

We hope to create a new vision for climate change engagement, and highlight the astronomy community's obligation to promote that vision. We hope that, by embedding climate change action in astronomy, we will be able to truly contribute to solving the most pressing problem of our time. We hope to engage, not only as astronomers, but as humans.